\documentclass[]{iopart}
\usepackage{graphicx}
\usepackage{amssymb}
\usepackage{cite}
\usepackage{iopams}
\usepackage[dvips]{color}
\usepackage{color}

\begin{document} 

\title[Coulomb interaction effects in graphene bilayers] 
{Coulomb interaction effects in graphene bilayers:\\
electron-hole pairing and plasmaron formation}
\author{Van-Nham Phan$^{1,2}$ and Holger Fehske$^1$}
 \address{$^1$ Institut f\"ur Physik, Ernst-Moritz-Arndt-Universit\"at 
Greifswald, 17487 Greifswald, Germany}
\address{$^2$ Institute of Physics, Vietnamese Academy of Science and Technology, PO Box 429, 10000 Hanoi, Vietnam}
\ead{fehske@physik.uni-greifswald.de} 
\begin{abstract} 
We report a theoretical study of the many-body effects of electron-electron 
interaction on the ground-state and spectral properties of double-layer 
graphene. Using a projector-based renormalization method we show that
if a finite voltage difference is applied between the graphene layers
electron-hole pairs can be formed and---at very low temperatures---an 
excitonic instability might emerge in a double-layer graphene structure. 
The single-particle spectral function near the Fermi surface exhibits 
a prominent quasiparticle peak, different from neutral (undoped) 
graphene bilayers. 
Away from the Fermi surface, we find that the charge carriers 
strongly interact with plasmons, thereby giving rise to a 
broad plasmaron peak in the angle-resolved photoemission spectrum. 
\end{abstract} 
\pacs{71.10.-w, 71.38.-k, 75.47.Gk, 71.70.Ej}
%\submitto{\NJP} 
%\maketitle
\section{Motivation}
Graphene based structures are most likely the building blocks
of future nanoelectronic devices. The reasons for this are 
manifold, but one may highlight that many of the exceptional
properties of this new class of low-dimensional materials, which arise
from the special form of energy spectrum near the so-called Dirac nodal points
and the related nontrivial topological structure of the wave function, 
can be easily modified by the application of external electric 
and magnetic fields, as well as by confining the sample 
geometry, by chemical doping, or by edge functionalization and 
substrate manipulation; for recent reviews see~\cite{CGPNG09,Pe10,AABZC10}.
  
Single-layer graphene (SLG)~\cite{NGMJKGDF05,ZTSK05}, a truly two-dimensional
(2D) crystal with remarkable mechanical properties, can be considered
as a gapless semiconductor with zero density of states at the Fermi level
and a linear energy dispersion close to the (inequivalent) 
corners of the Brillouin zone (${\rm K,\, K'}$ Dirac points). 
Thus the low-energy electrons are massless, chiral Dirac fermions. 
As a consequence any backscattering is suppressed and the charge 
carriers are almost insensitive to disorder 
and electron-electron interactions. Bilayer graphene 
(BLG)~\cite{NMMFKZJSG06,OBSHR06}, consisting of 
two AB-stacked, chemically bound graphene monolayers, 
is also a zero-gap semiconductor, if unbiased, 
but with a parabolic band dispersion. That is, in this material 
the low-energy electrons acquire a finite quasiparticle mass.  
Interestingly there exists another physical realization of 
a graphene-based double-layer structure that can be fabricated: 
In this system---usually referred to as graphene bilayer or 
double-layer graphene (DLG)---the two graphene sheets are separated 
by an dielectric so that the tunneling between the layers 
can be neglected~\cite{LS08b,ZJ08,Poea11,SG12}.

In BLG and DLG the charge carrier concentration 
can be controlled by simple application of 
a gate voltage~\cite{NGMJZDGF04,OBSHR06}. 
This electric field effect is fundamental for potential technological 
applications. For BLG even the band structure might be manipulated 
by an electric field, so that a gap between the 
valence and conduction bands varies between zero and mid-infrared 
energies~\cite{CNMPLNGGC07}.

BLG, besides being at present the only known semiconductor with a
tunable band gap, is intriguing also from a many-particle interaction 
physics perspective. While in SLG the interaction parameter is the 
fine-structure constant and Coulomb interaction effects are typically small, 
in BLG the interaction strength is alterable by changing the 
carrier density~\cite{SHD11}. For DLG, the ability to modify the 
carrier polaribility of an individual layer even implies that the interlayer 
Coulomb interaction can be converted from repulsive to attractive~\cite{ZJ08}. 
Since a gate bias across the two-layered graphene leads to a charge
imbalance in the layers, an attractive Coulomb interaction between the 
excess electrons and holes on opposite layers raises the possibility of 
formation of electron-hole bound states (excitons)~\cite{DH08}.   
Excitonic effects have been reported in the optical response of 
doped SLG~\cite{Ya11} and BLG~\cite{YDPCL09} as well. 
Most notably the BLG and DLG systems seem to be apromising candidate for 
electron-hole pair condensation~\cite{MBSM08,ZJ08,LS08b,NL10,HZW11}. 
In the weak coupling regime, exciton condensation is triggered 
by a Cooper-type instability of systems with 
occupied conduction-band states and empty valence-band states
inside identical Fermi surfaces~\cite{KK68,MBSM08,NL10}. The particle-hole 
symmetry of the Dirac equation obviously ensures a perfect nesting between
the electron Fermi surface and its hole counterpart in 
the $n$- and $p$-type layer of a biased DLG system~\cite{MBSM08}. 
A condensate of spatially separated electron-hole pairs then might occur when
the interlayer tunneling is negligible, but corresponding Coulomb 
interaction is not~\cite{ZJ08,LS08b}. Such bilayer exciton condensates typify 
as counterflow superfluids~\cite{EM04}. 

In characterizing many-body aspects of SLG, BLG and DLG, 
and particularly the   
electron-electron, electron-hole and electron-plasmon interaction 
effects, the single-particle spectral function $A(k,\omega)$ gives detailed 
information. Angle-resolved photoemission spectroscopy (ARPES) is a 
powerful probe of $A(k,\omega)$ in (quasi-) 2D single-, bi- and 
few-layer graphene because it achieves frequency and momentum 
resolution and thereby directly addresses the quasiparticle energy band 
properties~\cite{OBSHR06,BOSHR07,BSSHPAMR10}. So far the spectral
function of SLG and BLG has been calculated within the simple $G_0W$ and
random-phase approximation~\cite{PABBPM08,HD08b,HD08,SHD10,SHD11}, with a focus
on dynamic screening, Kohn anomaly, Friedel oscillations and plasmons.   

In this work, we employ the projector-based renormalization method (PRM) 
to determine the ground-state and spectral properties of an effective 
graphene bilayer  model 
describing the interaction and possibly pairing 
of electrons from the top bilayer with holes from the bottom layer [cf.
figure~\ref{fig:Intro}~(a)]. Our main interests here are the effects
of the Coulomb interaction regarding the formation of excitonic 
bound states and, for the doped system, the coupling
between the charge carriers and plasmons leading to composite 
``plasmaron'' quasiparticles. We show that graphene-based bilayers 
might allow for electron-hole pair condensation and exhibit, in their 
ARPES spectra, pronounced plasmaron signatures away from the Dirac point.
\section{Theoretical Approach}
\begin{figure}[hbt]
  \begin{center}
\includegraphics[angle = -0, width = \textwidth]{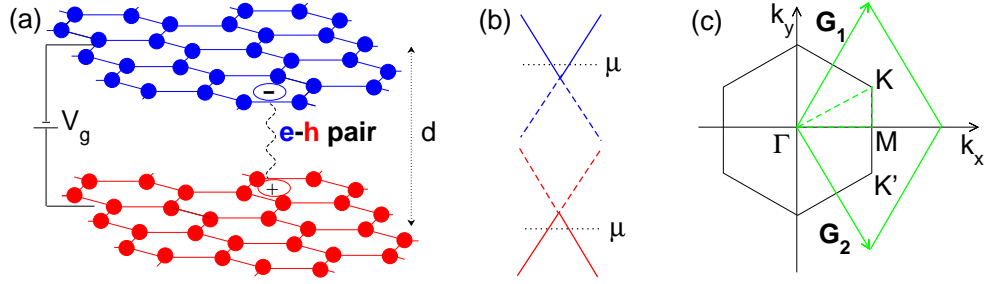}
\end{center}
  \caption{(a) Schematic of a graphene bilayer system with electron and hole
carriers induced by external gates forming excitonic bound states due to 
their Coulomb attraction. (b) The positions of the chemical potential
in the two graphene layers is adjusted by the voltage $V_{\rm {g}}$. (c) Brillouin 
zone of the graphene honeycomb lattice (black lines) and equivalent
Brillouin zone  (solid green lines) spanned by the inverse lattice vectors 
${\bf G}_1$ and $ {\bf G}_2$. Dirac cones are 
located at the ${\rm K}$ and ${\rm K^\prime}$ points.}
\label{fig:Intro}
\end{figure}
We consider two graphene sheets separated by a dielectric thickness
$d$, and assume a hexagonal stacking in which each sublattice in one 
layer is on top of the corresponding sublattice in the other
layer~\cite{ZJ08}. 
Let now $V_{\rm {g}}=eE_{\rm {ext}}d$ be a gate-induced potential
difference between the two layers. Then the external electric 
field $E_{\rm {ext}}$ outside the DLG system can be used to tune 
the chemical potential $\mu=V_{\rm {g}}/2$~\cite{MBSM08}: For a neutral 
graphene DLG system 
the Fermi level lies symmetrically in the conduction band of the upper 
($n$-type) layer and in the valence band of the lower ($p$-type) 
layer [see figure~\ref{fig:Intro}~(b)]. Within an effective two-band
description, the filled valence and empty conduction bands of the
$n$- respectively $p$-type layers can be neglected [dashed lines 
in panel~(b)], and the 
Hamiltonian of the graphene bilayer, 
in the absence of the Coulomb interaction, is 
\begin{equation} 
\mathcal{H}_0=g_{\rm s}\sum_{\mathbf{k}}\varepsilon^+_{\bf k}a^\dag_{\bf k}a^{}_{\bf k}+g_{\rm {s}}\sum_{\mathbf{k}}\varepsilon^-_{\bf k}b^\dag_{\bf k}b^{}_{\bf k}\,.
\label{H_0}
\end{equation}
Here $a^{\dag}_{\bf k}$ and $b^{\dag}_{\bf k}$ are the creation operators
of electron and hole quasiparticles with in-plane momentum ${\bf k}$
and band dispersions  
\begin{equation}
\varepsilon^\pm_{\bf k}=\pm\gamma_0 \left[1+4\cos^2\frac{k_y}{2}+4\cos\frac{k_y}{2}\cos\frac{\sqrt{3}}{2}k_x\right]^{1/2}\mp\mu\,, 
\label{banddis}
\end{equation}    
which for momenta close to the Dirac points become the linear
energy-momentum relations sketched by the blue and red straight 
lines in figure~\ref{fig:Intro}~(b), respectively. 
In equations~(\ref{H_0})--(\ref{banddis}), 
$g_{\rm {s}}=2$ allows for the spin degeneracy and  
$\gamma_0$ denotes the nearest-neighbor transfer amplitude
($\gamma_0\simeq 2.8$ eV~\cite{CGPNG09}).

In order to describe exciton formation, we take into account 
the interlayer coupling between electrons and holes,
\begin{equation} 
\mathcal{H}_{\rm e-h}=\frac{g_{\rm {s}}}{N}\sum_{{\bf k}_1{\bf k}_2,{\bf q}\ne 0}U_{{\bf k}_1{\bf k}_2{\bf q}} \;a^\dag_{{\bf k}_1+{\bf q}}a^{}
_{{\bf k}_1}b^\dag_{{\bf k}_2-{\bf q}}b^{}_{{\bf k}_2}\;,
\label{H}
\end{equation}
where
\begin{equation}
U_{{\bf k}_1{\bf k}_2{\bf q}}=\kappa \frac{e^{-d|{\bf q}|}}{|{\bf q}|}\cos\frac{\phi_1}{2}\cos\frac{\phi_2}{2}\quad\mbox{with}\quad\kappa=g_{\rm s}\frac{2\pi e^2}{\epsilon}\,.
\end{equation}
The Coulomb matrix element $U_{{\bf k}_1{\bf k}_2{\bf q}}$ 
contains besides the dielectric constant $\epsilon$, characterizing the 
embedding medium, a graphene-specific factor 
$\propto\cos\frac{\phi_1}{2}\cos\frac{\phi_2}{2}$, 
with scattering angles $\phi_i=\theta_{\mathbf{k}_i}-\theta_{\mathbf{k}_i+{\mathbf q}}$ where $\theta_{\mathbf k_i}=\textrm{atan}(k_{y}/k_{x})$~\cite{ZJ08}. 
At this point at least two remarks are in order. First, we have omitted the 
intralayer Coulomb repulsion between electrons or holes
that manifests itself simply by a screening of the 
interlayer Coulomb interaction. For BLG it has been shown
that the dynamically screened interlayer electron-hole interaction 
is attractive~\cite{LS08b}. Second the effective Coulomb attraction is 
long-ranged and depends in an involved way on the transmitted momentum. 
This is a major difference to the momentum-independent short-ranged 
Coulomb attraction in the extended Falicov-Kimball model,
recently studied in the context of exciton formation and condensation 
within the PRM approach~\cite{PBF10,PFB11}.
Excluding the ${\bf q}=0$ component (comprising the jellium background), 
the Hartree contributions to the one-particle energies vanish.
  
To proceed, we rewrite the total Hamiltonian, $\mathcal{H}=\mathcal{H}_0+
\mathcal{H}_{e-p}$,  in a normal-ordered form 
\begin{eqnarray}\label{HNO}
 \mathcal{H}=&\sum_{\mathbf{k}}\left[\varepsilon^+_{\bf k}a^\dag_{\bf k}a^{}_{\bf k}+\varepsilon^-_{\bf k}b^\dag_{\bf k}b^{}_{\bf k}+
(\Delta^{}_{\bf k}b^\dag_{\bf k}a^{}_{\bf k}+\textrm{H.c.})\right]\nonumber{}\\&\hspace*{2cm}+\frac{1}{N}\sum_{{\bf k}_1{\bf k}_2,{\bf q}\ne 0}U_{{\bf k}_1{\bf k}_2{\bf q}}:a^\dag_{{\bf k}_1+{\bf q}}a^{}_{{\bf k}_1}b^\dag_{{\bf k}_2-{\bf q}}b^{}_{{\bf k}_2}:\,
\end{eqnarray}
and look for a non-vanishing excitonic expectation value
\begin{equation}
\Delta_{\bf k}=-\frac{\kappa}{N}\sum_{\bf q}\frac{e^{-d|{\bf q}|}}{|{\bf q}|}
\,\frac{(1+\cos\phi)}{2}\,\langle a^\dag_{\bf{k+q}}b^{}_{\bf{k+q}}\rangle\,, 
\end{equation}
indicating a spontaneous symmetry breaking due to 
the pairing of electrons and holes. $\mathcal{H}$ is given in units 
of $g_{\rm s}$, and $\phi=\theta_{\mathbf{k}+{\mathbf q}}-\theta_{\mathbf{k}}$.

Now the single-particle spectral function, e.g. for the conduction band 
electrons, 
\begin{equation}\label{Aai}
A^{a}({\mathbf k},\omega)=-\frac{1}{\pi}\textrm{Im} \,G^a({\mathbf k},\omega) 
\end{equation}
with $G^a({\mathbf k},\omega)=\langle\langle a^{}_{\mathbf{k}};a^{\dagger}_{\mathbf{k}} \rangle\rangle^{}_{\mathcal{H}}$ being the Fourier transform of 
the retarded Green function ($\omega\rightarrow \omega+i0^{+}$),
can be computed by the PRM in close analogy 
to the calculation performed for the extended Falicov-Kimball model~\cite{PBF10}. We have
\begin{equation}\label{Ga}
G^a({\mathbf k},\omega)=\langle\langle \tilde{a}^{}_{\mathbf{k}};\tilde{a}^{\dagger}_{\mathbf{k}} \rangle\rangle^{}_{\tilde{\mathcal{H}}}\,,
\end{equation}
where ${\tilde{\mathcal H}}$ is a renormalized Hamiltonian
with all transition energies due to the interaction part of the 
Hamiltonian~(\ref{HNO}) are successively integrated out. 
After this renormalization procedure becomes complete, 
${\tilde{\mathcal H}}$ embodies 
a noninteracting Hamiltonian, but with a renormalized band structure, 
and therefore can be easily diagonalized by a Bogoliubov transformation
to yield
\begin{equation}\label{HReBogoliu}
{\tilde{\mathcal H}}
=\sum_{{\mathbf k}}\left(E^a_{{\mathbf k}}\hat{a}^\dagger_{{\mathbf k}}\hat{a}^{}_{{\mathbf k}}
+E^b_{{\mathbf k}}\hat{b}^\dagger_{{\mathbf k}}\hat{b}^{}_{{\mathbf k}}\right)\,,
\end{equation}
where
\begin{equation}
\label{qpe_c}
\fl E^{a/b}_{\mathbf k}=
-/+\frac{\textrm{sgn}(\tilde{\varepsilon}^-_{\mathbf k}-\tilde{\varepsilon}^+_{\mathbf k})}
{2}W_{\mathbf k}+\frac{\tilde{\varepsilon}^+_{\mathbf k}+\tilde{\varepsilon}^-_{\mathbf k}}{2}\quad\mbox{with}\quad
W_{\mathbf k}=\left[(\tilde{\varepsilon}^+_{\mathbf k}-\tilde{\varepsilon}^-_{\mathbf k})^2
+4|\tilde{\Delta}^{}_{\mathbf k}|^2\right]^{1/2}.
\end{equation}
The new fermionic operators satisfy  
$\hat{a}^{}_{\bf k}=u^{}_{\bf k}a^{}_{\bf k}+v^{}_{\bf k}b^{}_{\bf k}$ and 
$\hat{b}^{}_{\bf k}=v^{}_{\bf k}a^{}_{\bf k}-u^{}_{\bf k}b^{}_{\bf k}$,
with
$u^2_{\mathbf k}+v^2_{\mathbf k}=1$, 
$v^2_{\mathbf k}=[1-\textrm{sgn}(\tilde{\varepsilon}^-_{\mathbf k}
-\tilde{\varepsilon}^+_{\mathbf k})(\tilde{\varepsilon}^-_{\mathbf k}
-\tilde{\varepsilon}^+_{\mathbf k})/W_{\mathbf k}]$. Replacing the 
renormalized quasiparticle operators  
$\tilde{a}^\dagger_{\mathbf{k}}=\tilde{x}^{}_{\mathbf{k}}a^\dagger_{\mathbf{k}}+\frac{1}{N}\sum_{\mathbf{pq}}
\tilde{y}^{}_{\mathbf{kpq}}a^\dagger_{\mathbf{k+q}}
:b^\dagger_{\mathbf{p-q}}b^{}_{\mathbf{p}}:$
in equation~(\ref{Ga}), the spectral function is given to leading order
in $\tilde{\Delta}$ by
\begin{eqnarray}
\label{ImGcc}
\fl A^{a}&(\mathbf{k},\omega)=|\tilde{x}^{}_{\mathbf{k}}|^2\left[u^2_{{\mathbf k}}\delta(\omega-E^a_{{\mathbf k}})+v^2_{{\mathbf k}}\delta(\omega-E^b_{{\mathbf k}})\right]
%\nonumber{}\\ \fl &
+\frac{1}{N^2}\sum_\mathbf{pq}|\tilde{y}^{}_\mathbf{kpq}|^2\times\\\fl &\times\delta\left(\omega-(E^a_\mathbf{k+q}-E^b_{\mathbf p}+E^b_\mathbf{p-q})\right)
[\langle \hat{n}^a_\mathbf{k+q}\rangle(\langle \hat{n}^b_\mathbf{p-q}\rangle-
\langle \hat{n}^b_\mathbf{p}\rangle)+\langle \hat{n}^b_\mathbf{p}\rangle(1-\langle \hat{n}^b_\mathbf{p-q}\rangle)]\,.\nonumber
\end{eqnarray}
Note that the expectation values 
\begin{equation}
\hspace*{-1cm}\langle \hat{n}^a_\mathbf{k}\rangle =u^2_{\mathbf k}f(E^a_{\mathbf k})+v^2_{\mathbf k}f(E^b_{\mathbf k})\quad\mbox{and}\quad
\langle \hat{n}^b_\mathbf{k}\rangle=v^2_{\mathbf k}f(E^a_{\mathbf k})+u^2_{\mathbf k}f(E^b_{\mathbf k})\qquad
\label{ReNN}
\end{equation}
have to be calculated with the fully renormalized Hamiltonian 
${\tilde{\mathcal H}}$ as well, where $f(E^{a,b}_{\mathbf k})$ are 
Fermi functions. The spectral function for the valence-band holes, 
$A^{b}(\mathbf{k},\omega)$, is obtained in the exact same manner. 
\section{Results and Discussion}
The PRM approach outlined in the previous section must be numerically 
evaluated of course. In doing so, we work in the reciprocal space bounded by the
green lines in figure~\ref{fig:Intro}~(c), on a discrete set 
of $N=64\times 64$ points. Convergence of the whole renormalization scheme 
is assumed to be achieved if all quantities are determined with
a relative error less than 10$^{-5}$. Thereafter the spectral functions
are calculated using a Gaussian broadening in energy space of width 0.06.
In what follows, we choose $\gamma_0=1$ as unit of the energy, and measure
all energies relative to the Dirac-point chemical potential
of the balanced graphene bilayer. Furthermore, we fix the interlayer 
Coulomb coupling constant $\kappa=7$; such a value can be realized 
with, e.g., the commonly used $\rm SiO_2$ substrate/dielectric 
($\epsilon\simeq 4$)~\cite{LS08b}.   
\begin{figure}[htb]
    \begin{center}
      \includegraphics[angle = -0, width = 0.6\textwidth]{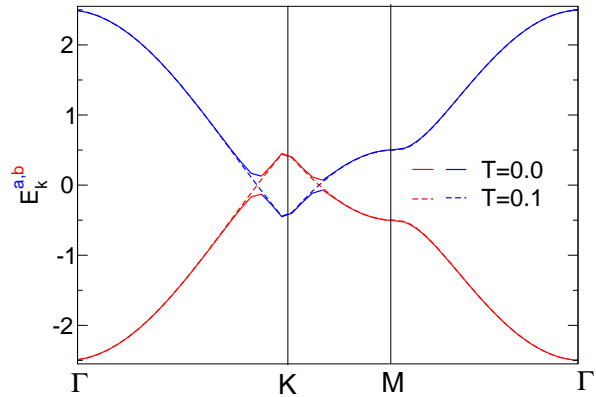}
\end{center}
\caption{Renormalized band dispersions, $E^{a,b}_{\mathbf{k}}$, along the 
high-symmetry directions of the Brillouin zone of graphene,  
indicated in figure~\ref{fig:Intro} by the dashed green lines. 
Results are given for an external field $E_{\rm{ext}}=1$ 
at zero temperature (solid lines) and at $T=0.1>T_{\rm {SF}}$ (dashed lines).}
\label{fig:EkT}
\end{figure}

Figure~\ref{fig:EkT} shows how the band structure of a suchlike interacting 
DLG system changes, if the electric field $E_{\rm{ext}}=1$ is applied.   
Apparently a pronounced excitonic gap appears at zero (low) temperatures.\footnote{Note that the natural stacking in graphite is the Bernal stacking
which merely leads to a insignificantly smaller excitonic gap 
however, see figure~2 in~\cite{ZJ08}.} 
The gap originates from electron-hole pair (exciton) condensation,
driven by the attractive Coulomb interaction over the full bandwidth.
In the result of this excitonic instability a macroscopically coherent 
quantum state emerges which, since electrons and holes are spatially 
separated in a graphene bilayer, differs in nature from the 
excitonic insulator phase in bulk 2D and 3D 
systems~\cite{KK65,HR68b,BSW91,BF06,IPBBF08,MSGMDCMBTBA10,PFB11,SEO11,ZIBF12}. 
As can be seen from figure~\ref{fig:EkT}, the renormalized quasiparticle 
bands $E_{\bf k}^{a,b}$ possess a nearly perfect particle-hole symmetry. 
The gap disappears at higher temperatures when interlayer 
quantum coherence is destroyed by thermal fluctuations. For a strictly 
2D system, of course, the critical temperature for exciton 
condensation would be zero, but the superfluid properties should 
survive for temperatures smaller than the Kosterlitz-Thouless 
transition temperature~\cite{ZJ08,MBSM08}. Therefore, for the DLG system,
our finite-temperature results will be valid for $T\leq T_{\rm {KT}}$ at least.

The zero- (low-) temperature exciton condensate is characterized 
by a non-vanishing expectation value 
$\delta_{\bf k}=\langle a^\dag_{\bf k}b^{}_{\bf k}\rangle$.
Since the interlayer Coulomb interaction between electrons and holes 
is attractive for all momenta and we exclusively consider a uniform 
exciton condensate in DLG, $\delta=(1/N)\sum_{\bf k}\langle a^\dag_{\bf k}b^{}_{\bf k}\rangle$ can be taken as an adequate order parameter for the electron-hole 
pairing.
\begin{figure}[b]
    \begin{center}
\includegraphics[angle = -90, width = 0.46\textwidth]{D_TEii.eps}          \hspace{0.1\linewidth}   
\includegraphics[angle = -90, width = 0.41\textwidth]{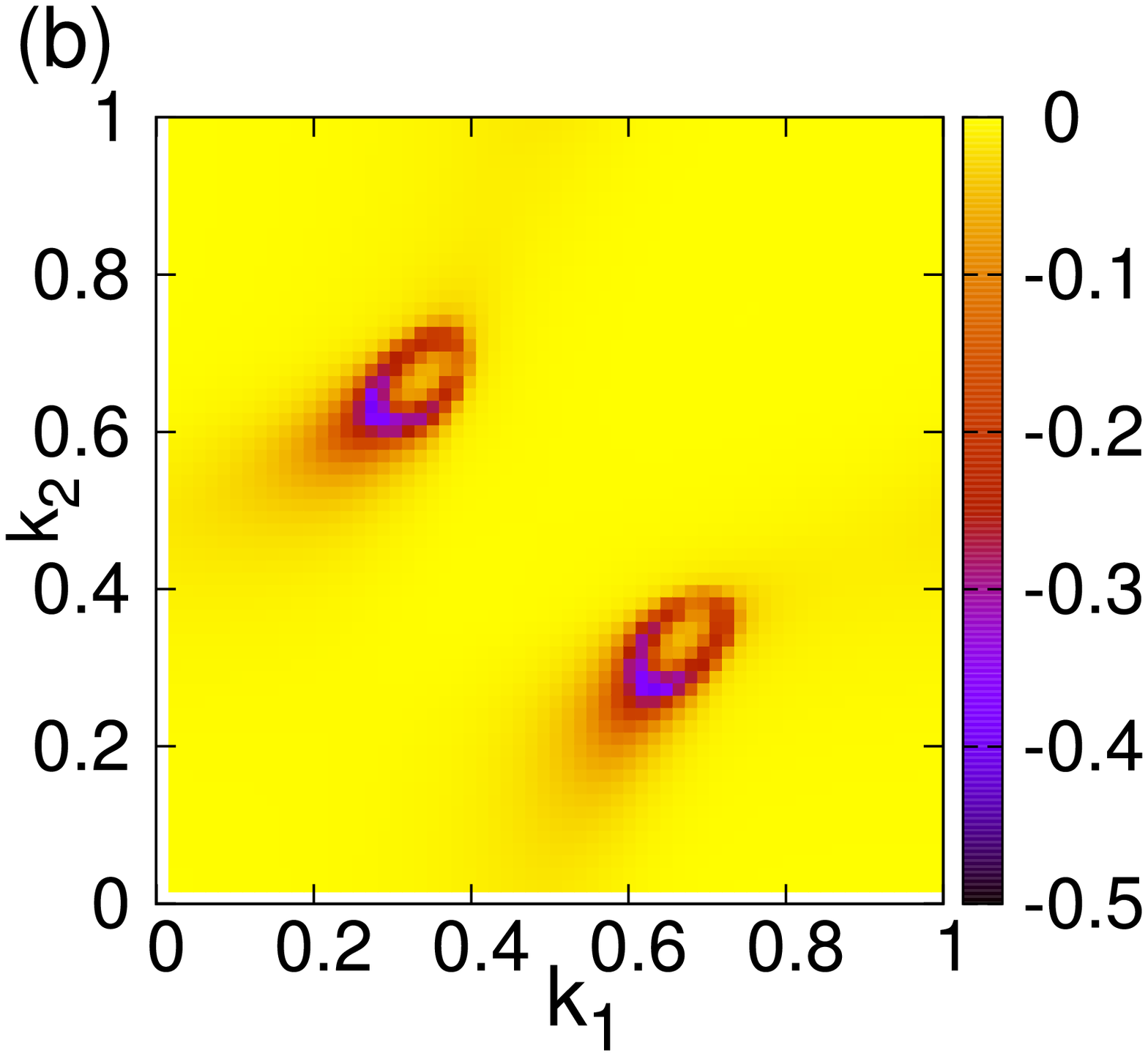}\\[0.05cm]
\hspace{0.038\linewidth}\includegraphics[angle = -90, width = 0.41\textwidth]{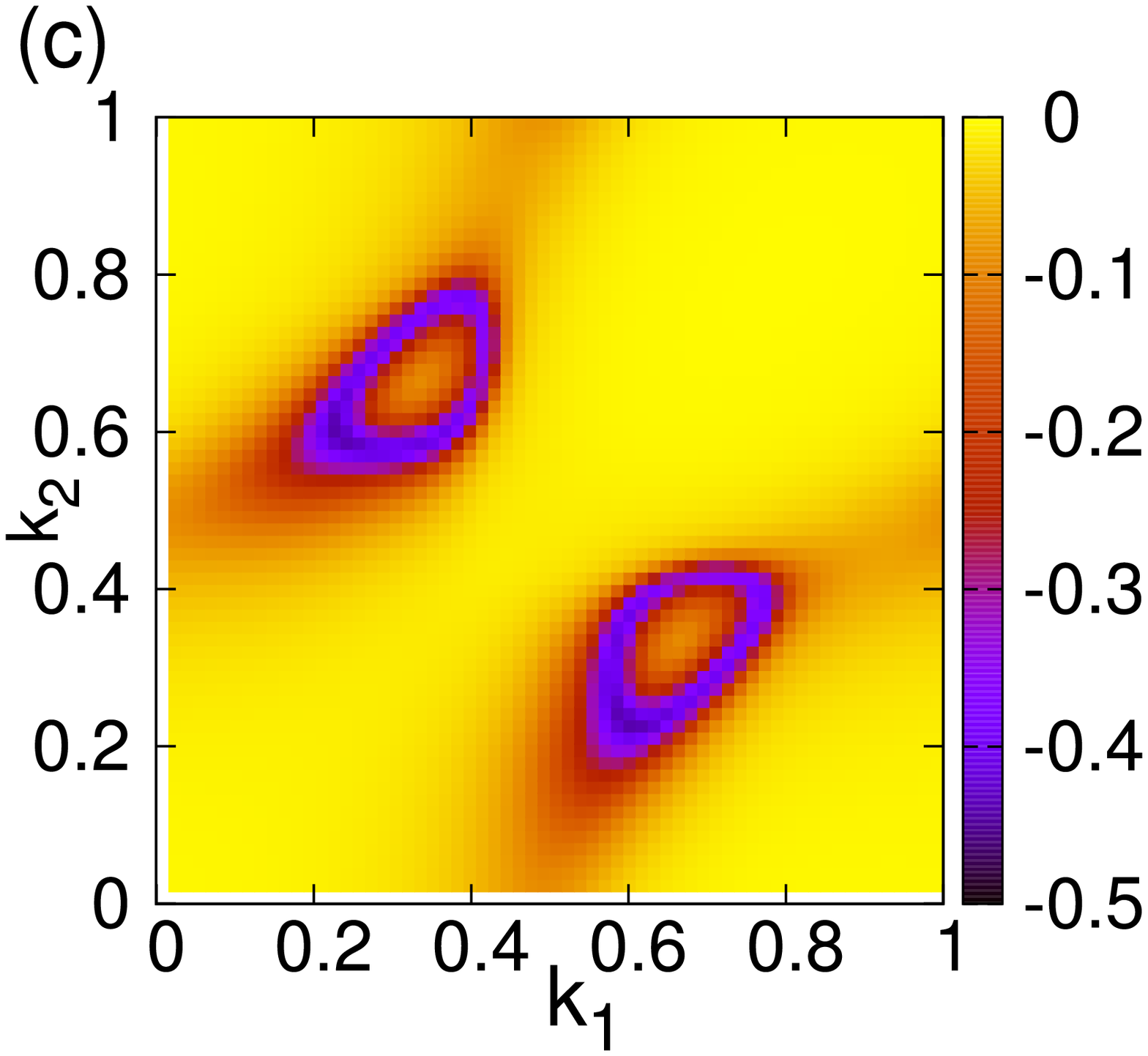}
\hspace{0.112\linewidth}  \includegraphics[angle = -90, width = 0.41\textwidth]{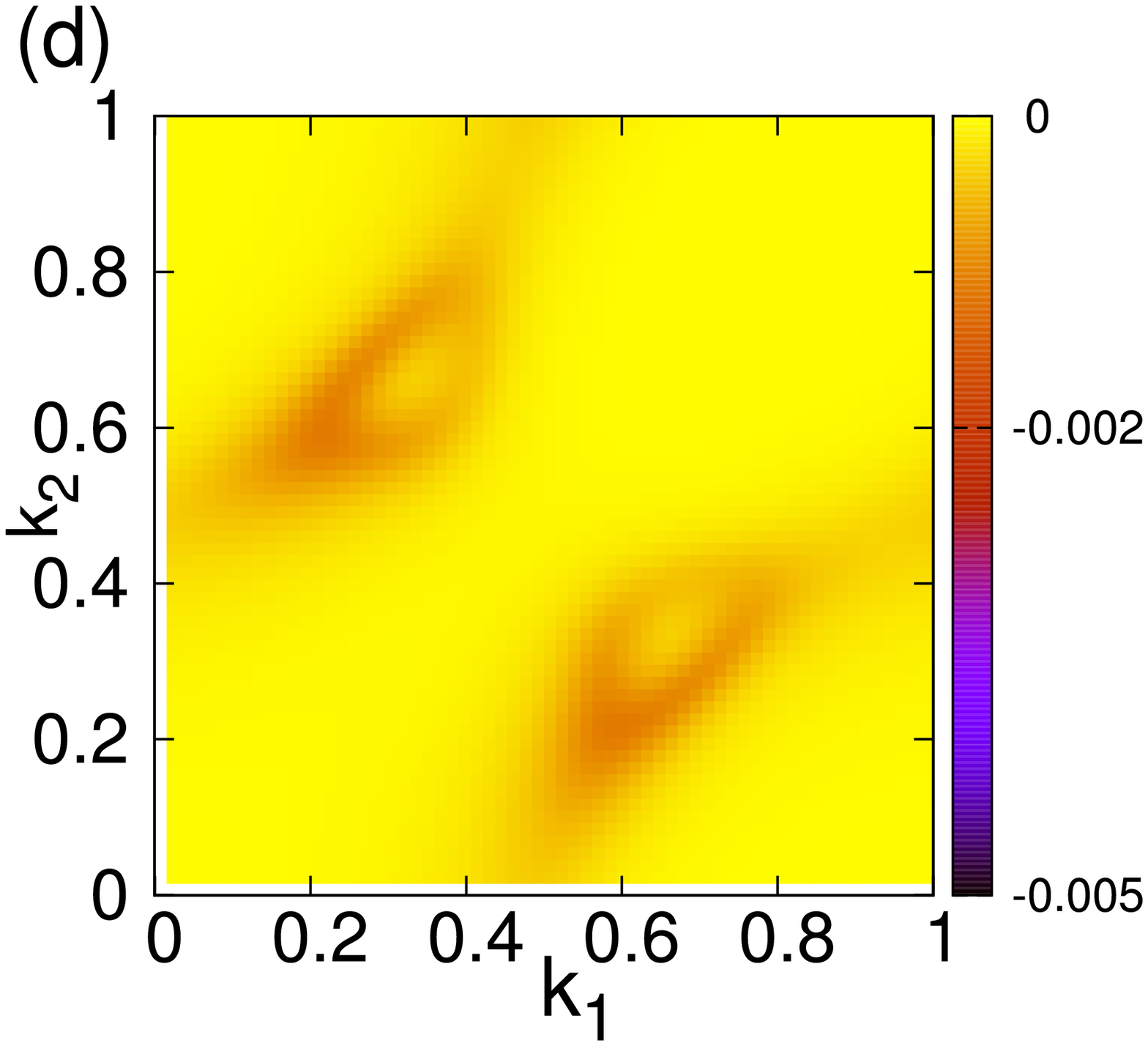}                  
    \end{center}
\caption{(a) Order parameter $\delta=(1/N)\sum_{\bf k}\langle a^\dag_{\bf k}b^{}_{\bf k}\rangle$ as a function of temperature $T$ for an 
external electric field $E_{\rm{ext}}=1$ (black circles) respectively as 
a function of $E_{\rm{ext}}$ at $T=0$ (red squares). The other panels show 
the magnitude of the expectation value, 
$\delta_{\bf k}=\langle a^\dag_{\bf k}b^{}_{\bf k}\rangle$ 
with ${\bf k}=k_1{\bf G}_1+k_2{\bf G}_2$, in reciprocal space 
by an intensity plot for $E_{\rm{ext}}=0.6$, $T=0$ [panel (b)], 
$E_{\rm{ext}}=1$, $T=0$ [panel (c)], and $E_{\rm{ext}}=1$, $T=0.1$ 
[panel (d); note the different intensity-color coding].}
\label{fig:Dk-Vg}
\end{figure}
This quantity is plotted in figure~\ref{fig:Dk-Vg}~(a) against the magnitude
of the electric field at zero temperature (red squares). It shows
that above the threshold value $E_{\rm{ext}}\simeq 0.4$ 
the excitonic condensate becomes more 
robust as more charge carriers (electrons and holes) are induced
into the layers by increasing the bias between them. Obviously, a  
larger valence and conduction band overlap amplifies the Cooper-like 
excitonic instability in the perfectly nested DLG system.
Fixing $E_{\rm{ext}}=1$, the order parameter is suppressed as 
the temperature increases (black circles), and vanishes 
when the electron-hole coherence is  completely lost in the 
high-temperature phase. As usually happens, the 
critical temperature itself is overestimated by theoretical 
approaches that comprise the electronic correlations and 
thermal fluctuations in an approximative way ($T_c\sim 0.1$~eV 
for $E_{\rm{ext}}=1$, see figure~\ref{fig:Dk-Vg}~(a)). We note, 
however, that $T_c$ is expected to be much larger than those of 
typical superconductors because exciton condensation is driven 
by the Coulomb interactions over the full bandwidth~\cite{MBSM08}. 
Panels~(b)--(d) yield detailed information 
on the variation of the the order parameter function $\delta_{\bf k}$ in 
the DLG Brillouin zone (here the diagonal belongs to 
the $\overline{\Gamma {\rm M}}$ direction). Raising the external 
electric field at zero temperature, exciton formation and condensation 
can set in if the conduction band of the upper layer enters 
the valence band of the lower layer. This takes place first 
at the  Dirac points. If the electric field is enhanced further,
electron and hole bands strongly hybridize along a crater-rim shaped 
contour around ${\rm K},\; {\rm K^\prime}$, thereby establishing electron-hole
coherence in this region [see panels (b) and (c)]. Above a critical 
temperature $T>T_{\rm {SF}}$ this signature is washed out and  
$\delta_{\bf k}$ becomes basically zero anywhere.

To gain further insight about the Coulomb interaction effects in DLG we
now analyze the PRM results obtained for the total wave-vector and frequency 
resolved single-particle spectral function $A({\bf k},\omega)= A^a({\bf k},\omega)+A^b({\bf k},\omega)$. Depending on the energy $\omega$, 
$A({\bf k},\omega)$ measures the probability of finding an 
electron (hole) with momentum ${\bf k}$ in the system and 
is directly accessible by photoemission (inverse photoemission) 
experiments. Thereby it gives valuable information about the 
quasiparticle properties, particulary with regard to the quasiparticle 
dynamics.

\begin{figure}[htb]
    \begin{center}
      \includegraphics[angle = -90, width = 0.415\textwidth]{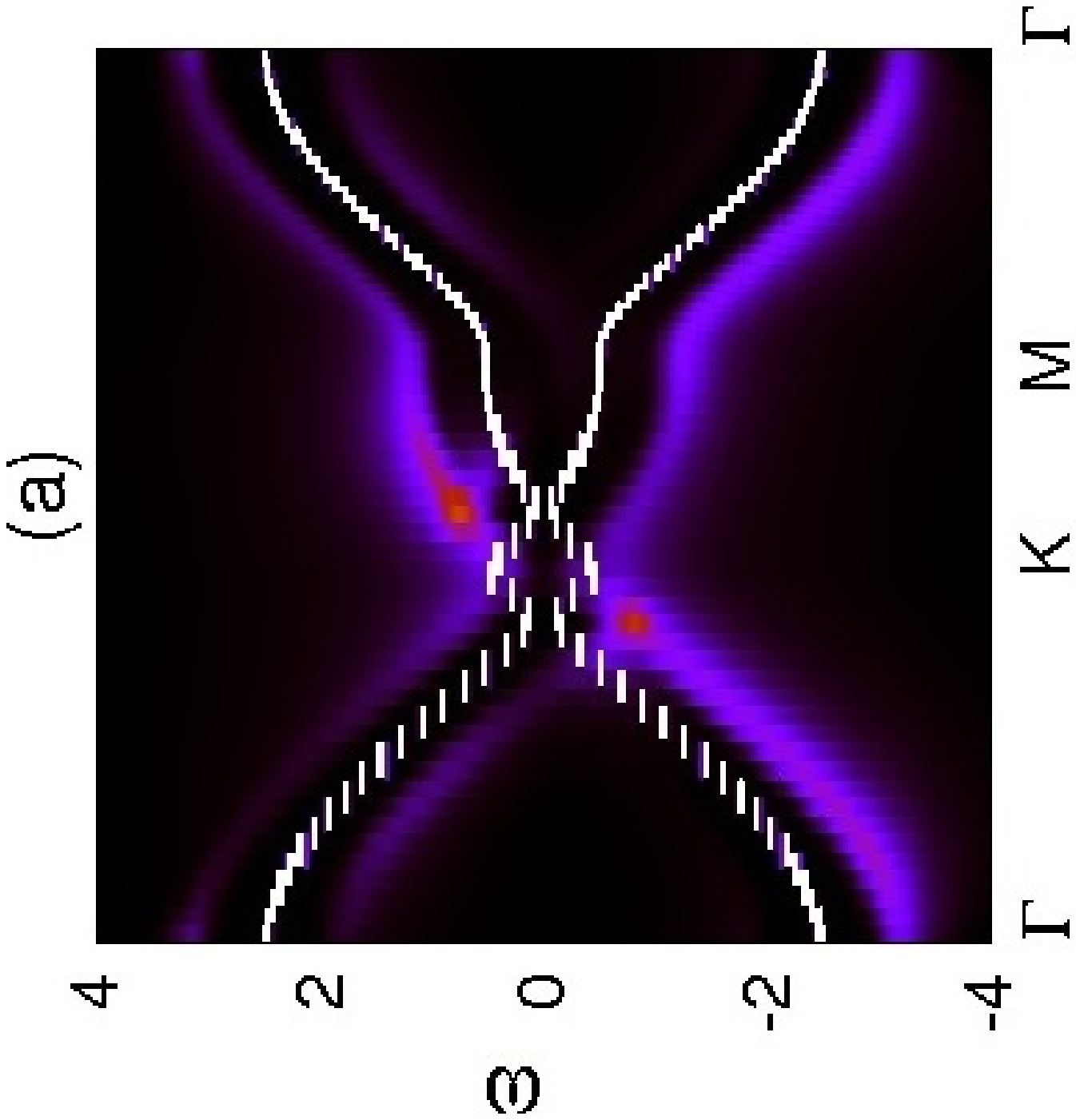} \hspace{0.05\linewidth}         
      \includegraphics[angle = -90, width = 0.47\textwidth]{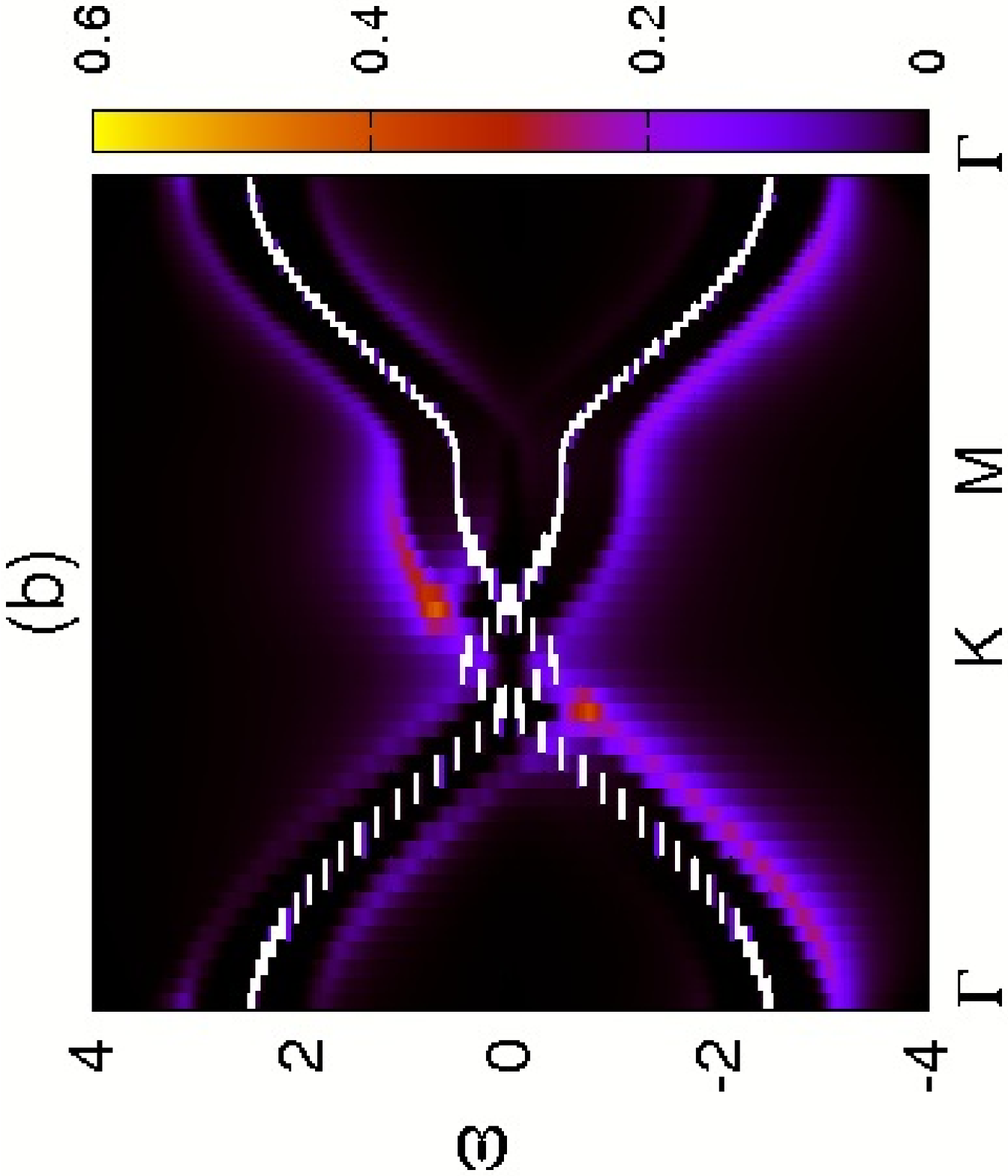}\\[0.3cm]
      \includegraphics[angle = -90, width = 0.415\textwidth]{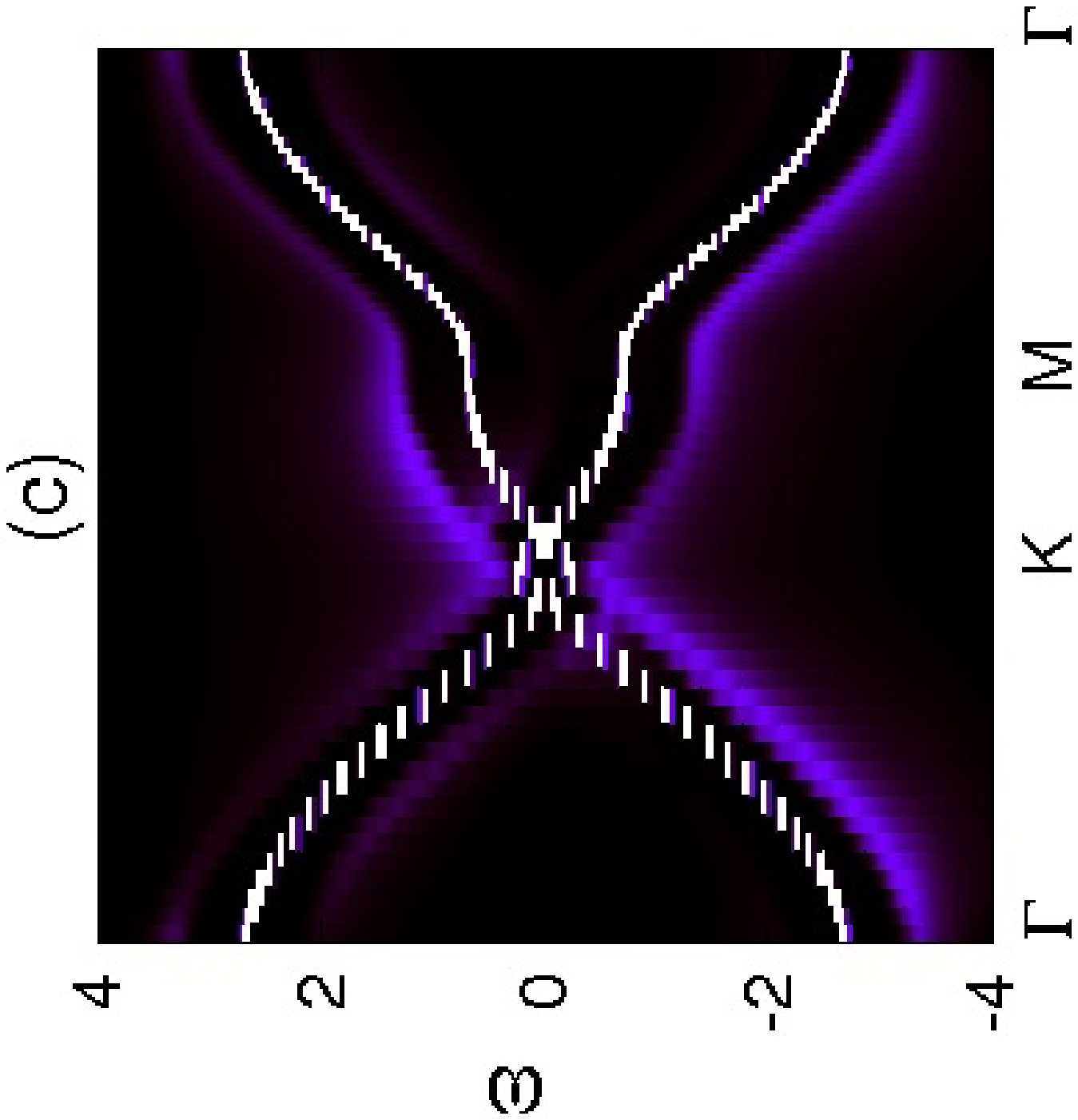}\hspace{0.05\linewidth}
      \includegraphics[angle = -90, width = 0.47\textwidth]{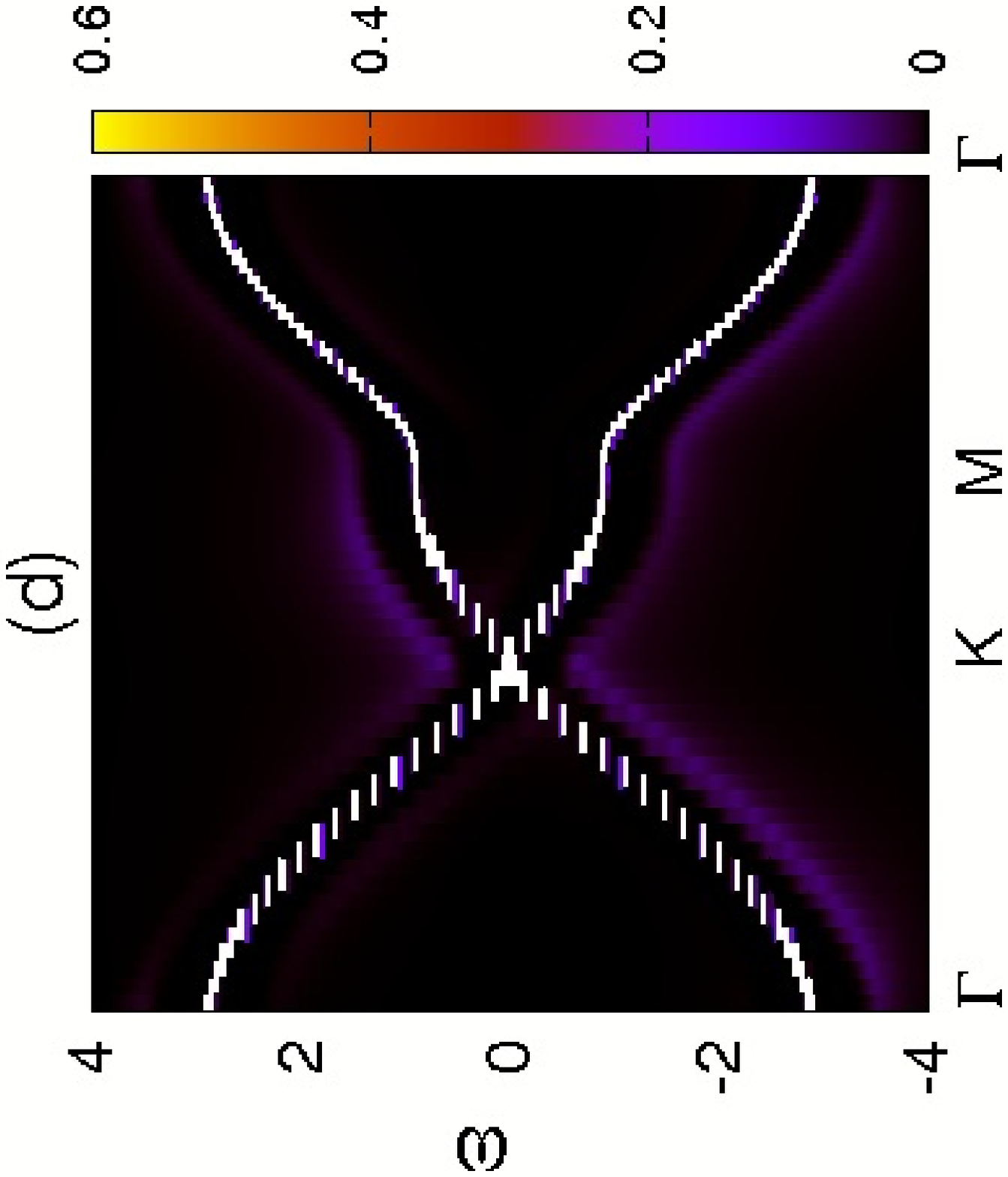}                  
    \end{center}
\caption{Intensity plots of the total wave-vector resolved single-particle
spectral function $A({\bf k},\omega)$ for $T=0$ and 
$E_{\rm{ext}}=1$ [panel (a)], respectively $T=0.1$
and  $E_{\rm{ext}}=1$ (b), 0.6 (c), and 0.2 (d).}
\label{fig:Skw-Vg}
\end{figure}

A key question in this respect is whether the non-Fermi-liquid behavior
observed for neutral undoped BLG/DLG at the Dirac point will survive
in the presence of any doping. While Coulomb interactions self-evidently 
play no role for a pure zero-density Fermi system, i.e. will not affect this  
non-Fermi-liquid fix point, they become important at any finite 
carrier density (induced, e.g., by applying a gate voltage), or if spatially 
density fluctuations are present in real DLG samples. Using the random-phase 
and parabolic band approximations, it has been shown quite recently, 
that BLG constitutes a Fermi liquid for doped cases, 
similar to the more standard 2D semiconductor-based 
electron gas~\cite{SHD11}. If so, the Coulomb 
interaction should cause noticeable correlation effects.

Figure~\ref{fig:Skw-Vg} gives the intensity of the single-particle 
spectral function, subject to $\omega$ along the high-symmetry directions of
the Brillouin zone, for several characteristic situations. 
Panel (a) displays the behavior of $A({\bf k},\omega)$ when an exciton 
condensate is realized at $T=0$. We find the location of the 
quasiparticle bands in accordance with the results of figure~\ref{fig:EkT}. 
These signatures have large spectral weight. Accordingly the 
effective-mass renormalization in DLG is expected to be rather weak, in contrast
to the large mass renormalization in a normal 2D electron gas~\cite{SHD11}. 
The single-particle excitation gap, located in the vicinity of the 
K~point, disappears for $T=0.1$ [panel~(b)]. That is, for $T>T_{\rm {SF}}$ 
we have gapless excitations only. Panels (c) and (d) show how the 
overlap of the predominantly electron ($\omega>0$) and hole ($\omega<0$) quasiparticle bands
shrinks as the electric field is reduced at fixed temperature ($T=0.1$).

Besides the quasiparticle band signal discussed so far, we 
observe---especially in panels (a)-(c)---another signature
that has less spectral weight, but is nevertheless well pronounced. 
It can be attributed to a plasmaron~\cite{Lu67}.
In an electron gas, the long-range Coulomb interaction manifests itself through
plasmonic collective charge density oscillations. These plasmons
can in turn interact with the charge carriers, with the result
that a new composite particle called plasmaron is formed~\cite{Lu67}.     
Plasmaron quasiparticles have been predicted to occur in 
SLG~\cite{HD08a,PABBPM08,LCN11} and BLG~\cite{SHD11}, and recently were 
observed in the former system by ARPES~\cite{BSSHPAMR10}. 
Our, regarding the treatment of interaction and lattice
structure effects more involved PRM calculation corroborates 
this prediction for graphene bilayers. As can be seen from
figure~\ref{fig:Skw-Vg}, satellite bands, arising from the dressing of the
injected particle (electron or hole) by plasmon modes, appear 
away from the Fermi momentum. The plasmaron shows up at greater
(smaller) frequency than the quasiparticle because of the extra 
energy cost for the binding of the electron (hole) with the plasmon.
The higher intensity of the plasmaron in the lower branch of the 
spectrum can be attributed to the fact that the plasmon oscillations 
are more pronounced in the $p$-type (few-hole) layer. The plasmaron
signature weakens, of course, at very high doping level. 
Increasing the temperature, the plasmaron signal is 
slightly enhanced initially (since the electron-hole pairing is 
suppressed), but finally is smeared out when the temperature 
achieves the order of magnitude of the particle-plasmon binding energy. 
For small gate voltages [panel (d)], the plasmaron dispersion resembles 
that of SLG~\cite{HD08a,LCN11}, simply because the band overlap is small 
or even absent and the quasiparticles in the $n$- and $p$-type branches 
become uncorrelated.

\begin{figure}[htb]
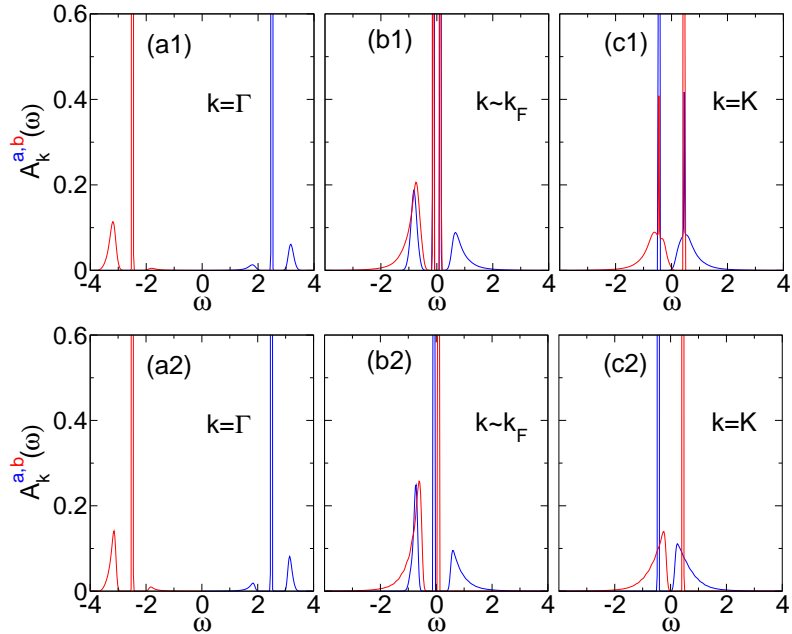

    \begin{center}
      \includegraphics[angle = -0, width = 0.8\textwidth]{Awk_G-K-kF_T00E10.eps}\\[0.1cm]
      \includegraphics[angle = -0, width = 0.8\textwidth]{Awk_G-K-kF_T01E10.eps}            
    \end{center}
\caption{Single-particle spectral functions of conduction-band electrons
$A^a({\bf k},\omega)$ (blue lines) 
and valence-band holes $A^b({\bf k},\omega)$ (red lines)
plotted against $\omega$ for characteristic momenta
${\bf k}=\Gamma$ [left-hand (a)-panels],  ${\bf k}\sim {\bf k}_{\rm {F}}$ [middle 
(b)-panels],
and  ${\bf k}=K$ [right-hand (c)-panels] at $T=0.0$, $E_{\rm {ext}}=1.0$ 
(upper row 1) and $T=0.1$, $E_{\rm {ext}}=1.0$ (lower row 2).}
\label{fig:Akw}
\end{figure}
Figure~\ref{fig:Akw} finally presents the frequency dependence of the
partial electron and hole spectral functions at
selected ${\bf k}$ points. It shows that the quasiparticles far away
from the Fermi point are predominantly hole-like (red curves) or electron-like 
(blue curves) at both zero and finite temperature; cf. the (a) panels. 
This holds as well as for the corresponding plasmaron signatures.  
For ${\bf k}\simeq {\bf k}_{\rm {F}}$  (Fermi wave vector), the 
single-particle excitations are gapful (gapless) for 
$T=0$ ($T>T_{\rm {SF}}$). Most notably, we observe a 
strong mixing of electron and hole bands, which also becomes 
apparent for the plasmarons. 
Moving to the K-point, this mixing is lost [cf. (c)-panels], and the plasmaron
bump appears well below (above) the corresponding 
electron (hole) quasiparticle peak.

\section{Concluding remarks}
Analyzing by means of a projector-based renormalization method
Coulomb interaction effects within a generic graphene bilayer model,
we suppose that electron-hole pair formation and also condensation 
might appear in a DLG system at least at zero temperature, 
if charge imbalance between the layers is induced by an external 
electric field. Thereby the condensation of the spatially separated,
but bounded electrons and holes turns out to be a robust phenomenon 
that will not require a precise alignment of the graphene sheets 
embedded in the dielectric~\cite{ZJ08,MBSM08}. 
The excitonic instability is of Cooper (BCS) type, and manifests
itself by an arising interlayer coherence between the electron and hole
quasiparticles residing inside nearly perfectly nested Fermi surfaces.
The condensed superfluid low-temperature state in DLG 
differs from the excitonic insulator state observed 
in (intermediate valent) rare-earth~\cite{BSW91} 
or transition-metal systems~\cite{MSGMDCMBTBA10}. 
The exciting BCS-BEC crossover upon an increase of the coupling 
strength~\cite{BF06,SEO11,ZIBF12},
discussed for the latter materials as well as for standard semiconductor-based 
coupled quantum well structures, seems to be achievable in DLG 
in the presence of a perpendicular magnetic field only, when localized 
magnetoexcitons will form (see~\cite{LS08b} and references therein). 
The pronounced quasiparticle peak in the single-particle spectra 
indicates that the electrons (holes) in doped DLG behave more like in a 
usual Fermi liquid. The ARPES spectra, in addition, feature a plasmaron peak,
arising from the coupling of the charge carriers to collective 
density oscillations. 
Quite recently, the optical response of DLG for linearly 
polarized evanescent modes has been analyzed, showing exponential
amplification for transverse polarization, which yields evidence
for transverse plasmons, uniquely related to the chirality of the 
electronic carriers of graphene~\cite{SG12}. These effects  
might be of importance for the functionality of new 
plasmonic devices merging photonics and electronics~\cite{Ra08,RR08}.

\ack We thank K. W. Becker, F. X. Bronold, D. Ihle, R. Nandkishore and B. Zenker 
for stimulating discussions. This work was supported by the DFG 
through SFB 652 and SPP 1459. V.-N. P. was also 
financially supported by the National Foundation for Science and 
Technology Development (NAFOSTED) of Vietnam.
\section*{References}
%%%%\bibliography{articleBiGraphene,articlePRM,articleEFKM,articleEPAM}
%\bibliography{ref.bib}
%\bibliographystyle{iopart-num}
\providecommand{\newblock}{}

\end{document}